\documentclass[sigconf,nonacm]{acmart}%% Fonts used in the template cannot be substituted; margin 
\AtBeginDocument{%
  \providecommand\BibTeX{{%
    \normalfont B\kern-0.5em{\scshape i\kern-0.25em b}\kern-0.8em\TeX}}}

\begin{document}

\title{Spatial Orchestra: \linebreak Locomotion Music Instruments through Spatial Exploration}

\author{You-Jin Kim}
\email{yujnkm@ucsb.edu}
\orcid{0000-0003-0903-8999}
\affiliation{%
  \institution{University of California Santa Barbara}
  \city{Santa Barbara}
  \state{CA}
  \country{USA}
}

\author{Myungin Lee}
\email{myungin@umd.edu}
\orcid{0000-0002-4202-4364}
\affiliation{%
  \institution{University of Maryland, College Park}
  \city{College Park}
  \state{MD}
  \country{USA}
}

\author{Marko Peljhan}
\email{peljhan@ucsb.edu}
\orcid{0009-0000-2186-9256}
\affiliation{%
  \institution{University of California Santa Barbara}
  \city{Santa Barbara}
  \state{CA}
  \country{USA}
}

\author{JoAnn Kuchera-Morin}
\email{jkm@create.ucsb.edu}
\orcid{0000-0002-3542-567X}
\affiliation{%
  \institution{University of California Santa Barbara}
  \city{Santa Barbara}
  \state{CA}
  \country{USA}
}

\author{Tobias Höllerer}
\email{holl@cs.ucsb.edu}
\orcid{0000-0002-6240-0291}
\affiliation{%
  \institution{University of California Santa Barbara}
  \city{Santa Barbara}
  \state{CA}
  \country{USA}
}

\renewcommand{\shortauthors}{Kim et al.}

\begin{abstract}
Spatial Orchestra demonstrates how easy it is to play musical instruments using basic input like natural locomotion, which is accessible to most. Unlike many musical instruments, our work allows individuals of all skill levels to effortlessly create music by walking into virtual bubbles. Our Augmented Reality experience involves interacting with ever-shifting sound bubbles that the user engages with by stepping into color-coded bubbles within the assigned area using a standalone AR headset. Each bubble corresponds to a cello note, and omits sound from the center of the bubble, and lets the user hear and express in spatial audio, effectively transforming participants into musicians. This interactive element enables users to explore the intersection of spatial awareness, musical rhythm that extends to bodily expression through playful movements and dance-like gestures within the bubble-filled environment. This unique experience illuminates the intricate relationship between spatial awareness and the art of musical performance.

\smallskip
\textit{ This is a preprint version of this article. The final version of this paper can be found in the Extended Abstracts (Interacticity) of ACM CHI 2024. For citation, please refer to the published version.}
\textit{This work was initially made available on the author's personal website [yujnkm.com] on April, 2024, and was subsequently uploaded to arXiv for broader accessibility.}

\end{abstract}




\begin{teaserfigure}
  \includegraphics[width=\textwidth]{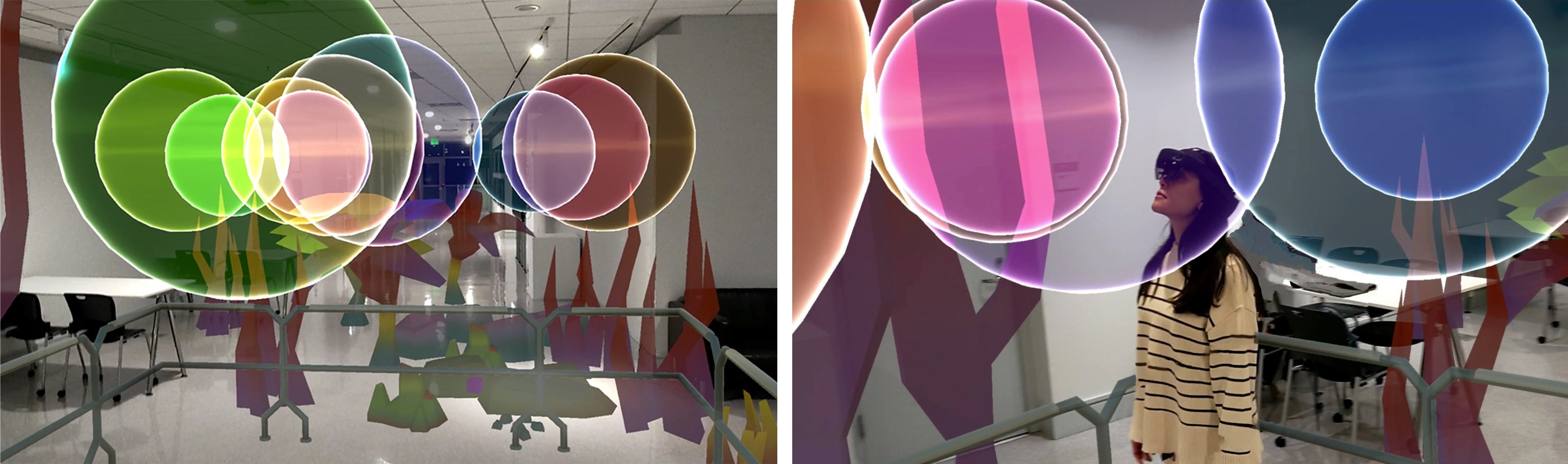}
  \caption{Spatial Orchestra. Left: From the point of view of the user experiencing Spatial Orchestra, where they are in the augmented stage, freely moving around and interacting with ten virtual bubbles that are ever shifting and simulate cello chords. Right: A user immersed in the augmented environment. }
  \Description{Figures illustrate view of the user experiencing Spatial Orchestra, where they are in the augmented stage, freely moving around and interacting with ten virtual bubbles that are ever shifting and simulate cello chords.}
  \label{fig:teaser}
\end{teaserfigure}


\maketitle

\section{Introduction}

\begin{figure*}[t]
 \includegraphics[width=\textwidth]{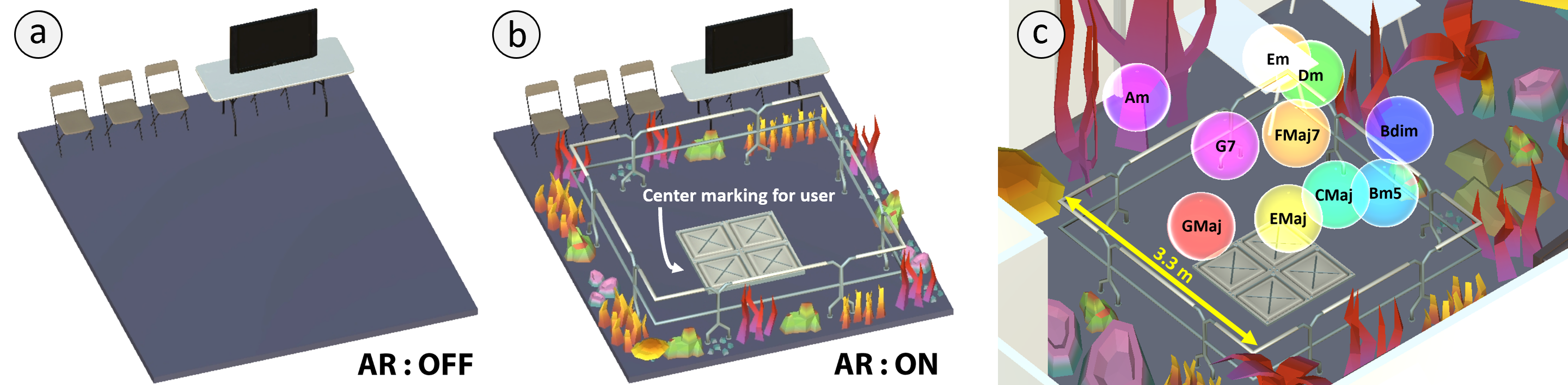}
 \caption{An aerial perspective of the stage arrangement that the user explores.  a. The physical space used for the Spatial Orchestra.  b. The fencing and virtual stage are all augmented.  c. There are ten augmented virtual bubbles set up in the area. Each bubble emits unique cello notes that are color coded to enable user interaction.}
 \label{fig:bubbles}
 \end{figure*}

In a world where not everyone has had the opportunity to acquire the skill of playing a musical instrument, the art of musical expression remains elusive, akin to a second nature for a select few. Those who possess this ability often find themselves in awe of those who can skillfully wield musical instruments to convey their emotions. This universal yearning for musical expression resonates with many individuals.

Within this context, we identified a novel application that seeks to answer the question, "What if individuals could create music using actions that are inherently familiar and well-practiced, utilizing the muscles and movements they employ in their daily lives?" This led to the conceptualization of a musical instrument designed to be played entirely through locomotion and spatial awareness, engaging users in an intuitive interaction with gently drifting bubbles. Within this defined area, color-coded bubbles, each emitting a distinct cello note, float at the user's head level. User can freely move around the play-space, engaging with these bubbles to compose music by entering them. Users are encouraged to navigate to areas where multiple bubbles are present, as user can create harmonies of mixed cello notes. 

Spatial Orchestra demonstrates research that utilizes on natural locomotion within mixed reality in open spaces~\cite{cheok2002interactive, kim2023dynamic, kim2022investigating, kim2023reality}, focusing on spatial awareness facilitated by spatial audio~\cite{RiddershomBargum2023, Earnormous, DuoRhythmo, AuditoryImmersion} in recent years. Our system leverages the spatial awareness that we employ in our daily routines, eliminating the need for users to acquire traditional music notation or score-reading skills. Instead, users can engage with the floating bubbles in a manner analogous to real-life interactions. By merging spatial awareness with the act of navigating their environment, individuals can become adept at playing an augmented reality (AR) musical instrument right from the start, offering a unique opportunity for everyone to become proficient music player. Moreover, users can further refine their skills with practice and experience. This means the users can not only produce incidental sound but also can develop their virtuosity to play sophisticated music. As proven by recent work, there are physical input-based and predictive approaches to musical embodiment~\cite{FormFollowsSound, maes2014action, maes2016sensorimotor, DuoRhythmo}, our work, Spatial Orchestra fully engages this concept and utilizes musical embodiment and the role of prediction, designing interactions through user input and action-based effects on music perception.

% trigger&playback-based Composition. Granular synth

% An active locomotion increases the probability of hitting the bubbles which are mapped to dynamic notes. 
% Why locomotion is a good way to produce music?: Music perception is a dynamic process firmly rooted in the natural disposition of sounds and the human auditory and motor system. vice versa. [Action-based effects on music perception], [Sensorimotor Grounding of Musical Embodiment and the Role of Prediction: A Review] https://www.frontiersin.org/articles/10.3389/fpsyg.2013.01008/full
% Lee thesis: intrinsic correlation between our body and sound

% [2015] Form Follows Sound: Designing Interactions from Sonic Memories, CHI : good arguments

\begin{figure*}[t]
\centering \includegraphics[width=\textwidth]{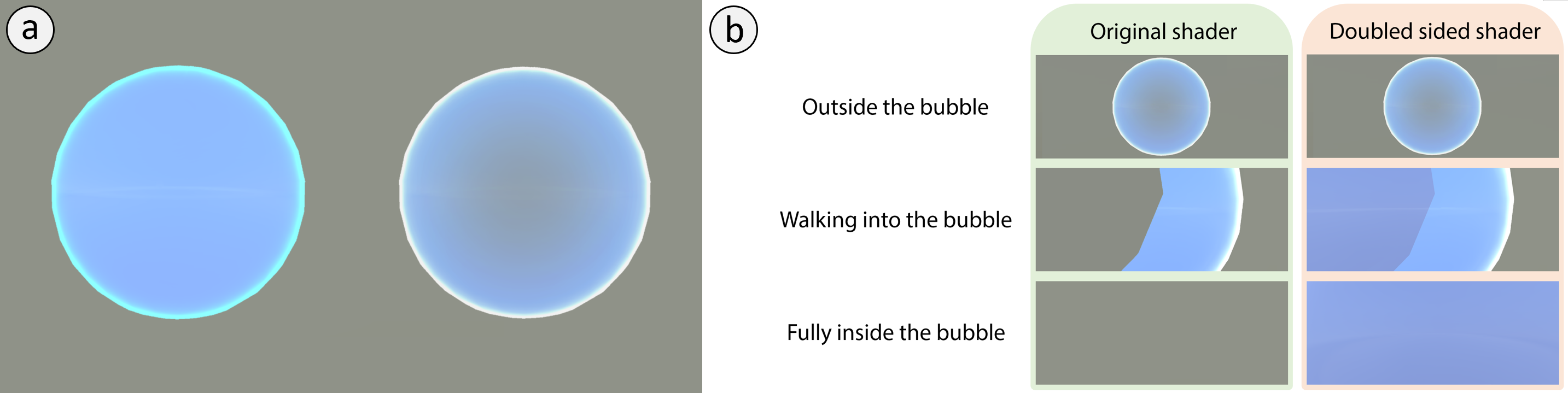}
 \caption{a. The bubble rendering model underwent improvements based on user feedback.  Left: The original bubble featured a color highlight at the edge, representing the bubble's assigned color. Right: The original bubble used had a color highlight of the bubble’s assigned color at the edge of the bubble.  Right: A new shader incorporated a white highlight for improved visibility of the bubble's boundaries. Adjustments were also made to the transparency and illumination levels, addressing user concerns about clarity when multiple bubbles are stacked and colors are mixed.  b. We addressed a critical issue with the original shader – users struggled to identify if they were inside the bubble. To solve this, we added a visual cue. Now, when users enter the bubble, the scene adopts a translucent color within the bubble's view. This was achieved through a double-sided shader, rendering both sides of the mesh.}
 \Description{Figure A illustrates the bubble rendering model underwent improvements based on user feedback.  Figure B illustrates a critical issue with the original shader.  Users struggled to identify if they were inside the bubble. To solve this, we added a visual cue. Now, when users enter the bubble, the scene adopts a translucent color within the bubble's view. This was achieved through a double-sided shader, rendering both sides of the mesh. } 
 \label{fig:boundary}
 \end{figure*} 

\section{Related Work}
Advances in digital medium facilitates new ways of musical expression. The electronic medium gave freedom from the acoustic and physical design of the instrument, allowing the development of wireless embodied interaction. While conventional instruments require years of training and mastery, this circumstance encouraged the researchers to propose unique and easy ways to expand the users' creativity and think outside the box.

\subsection{Digital Instruments}
Gehlhaar's work~\cite{soundspace91} uses an ultrasonic echolocation system to specify the locomotion of participants within the space, and the computer synthesizes the sound using the information. Morreale's installation \cite{Morreale14} invites the audience to compose music by moving in the space using the distance between the participants. Lee's work \cite{lee21} utilizes machine learning-based gestural recognition using smartphones' accelerometers and gyroscopes to interact with simulated physics within an immersive projection space. In Junior's work~\cite{Junior21}, non-experts create coherent music through graphical elements in a virtual environment.
 
This circumstance has allowed a proactive music experience for conventional music consumers through more familiar mediations, including new instruments and games, while engaging in compositional techniques~\cite{winter2005interactive}. Rasamimanana's work \cite{Rasamimanana12} mapped the sports ball's movement and status into rhythmic sounds. These types of musical games utilize the concept of the trigger and playback-based composition methods.  

\subsection{Musical Experience in Mixed Reality}
The advance of XR technology has revolutionized the way we interact with our surroundings, seamlessly blending the digital and real worlds. One intriguing area of research involves the experimentation of Spatial Audio in mixed reality.

Schlagowski et al. (2023) explored the fusion of Spatial Audio with hand motion-controlled interfaces in Virtual Reality, enabling users to collaboratively mix and play music in a Mixed Reality environment~\cite{schlagowski2023wish}. Another noteworthy project, DuoRhythmo by Riddershom and Bargum (2023), introduced a collaborative and accessible digital musical interface in mixed reality, focusing on designing a user-friendly experience~\cite{DuoRhythmo, RiddershomBargum2023}.

In the realm of education and entertainment, there have been efforts to leverage VR and Spatial Audio to enhance the way humans perceive and interact with games~\cite{Earnormous}. Additionally, projects like those enhancing auditory immersion in interactive VR environments demonstrate the diverse applications of Spatial Audio in immersive experiences~\cite{AuditoryImmersion}.

Turchet's research in 2021 highlighted a significant surge in the historical distribution of musical XR research over the past five years. The study aimed to define the Musical XR field by analyzing 260 research characteristics~\cite{Luca21}. Bilow (2022) proposed an Augmented Reality (AR) experience allowing participants to explore audiovisual elements through movements and to interact, using hand gestures~\cite{Bilbow22}. Furthermore, Wang (2022) conducted an empirical study comparing three audiovisual interface prototypes for head-mounted AR environments~\cite{wang22}.

These advancements collectively underscore the evolving landscape of XR technology, particularly in the integration of spatial audio for diverse applications ranging from collaborative music creation to educational experiences.

\section{Design of Spatial Orchestra}
In the virtual environment, to guide users to stay safely within the area, 3.3 m by 3.3 m space with a virtual fence. The fence is stylized realistically to differentiate it from other augmented objects in space. Ten bubbles, each measuring 80 cm in diameter, travel at a constant altitude level with the user's height, one meter every five seconds. Bubbles travel at random and bounce around the area while maintaining their altitude. Such behavior establishes a stochastic model like molecules in the closed space. 

Each bubble contains synthesized cello sounds that could be made from a single bow stroke. Sound omitted from the bubble's center can only be heard when the user's head is in the bubble. The colors of the ten bubbles represent the fundamental chords comprising: $[EMaj, Em, FMaj7, GMaj, G7, Am, Bdim, Bm5, Cmaj, Dm]$.
These are deliberately selected to guide participants in generating a musical structure that can fold and unfold over time through variant chord progressions. While the bubbles travel like molecules in space, the user’s vibrant motion will increase the probability of triggering the cello strokes. Through the integration of melodic framework with their spatial and visual perception, users can craft imaginative multimodal experiences. (Figure \ref{fig:bubbles}c). 

% generative composition and parametric constraints are discussed here. - ML

\begin{figure*}[t]
\centering \includegraphics[width=\textwidth]{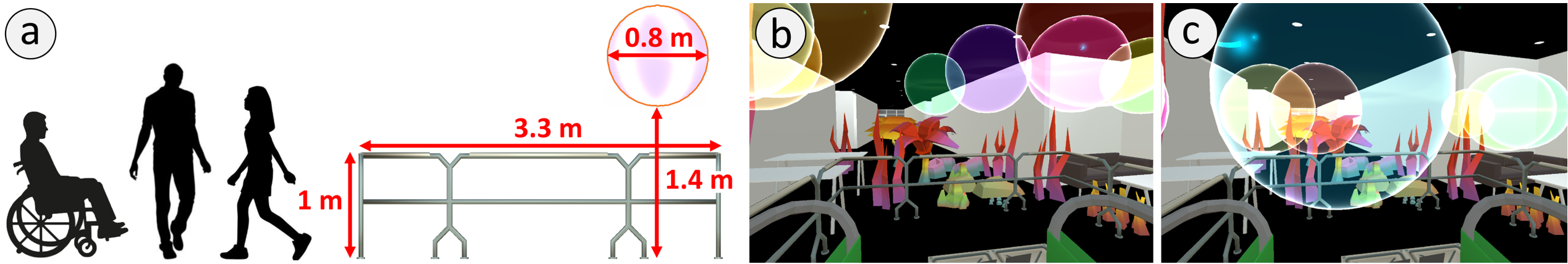}
 \caption{a. Compares the virtual bubbles' size to the virtual fence's height. Based on individual viewpoint heights, it may be challenging for some people to interact with bubbles directly.  b. View from the person in a wheelchair engaging with the bubbles.  c. We implemented an accessibility mode that allowed manual adjustment to the height of the bubbles. As you can see, the bubbles appear at the correct height when viewed from a wheelchair.}
 \Description{Figure A compares the virtual bubbles' size to the virtual fence's height. Based on individual viewpoint heights, it may be challenging for some people to interact with bubbles directly.  Figure B illustrates a view from the person in a wheelchair engaging with the bubbles.  Figure C shows how we implemented an accessibility mode that allowed manual adjustment to the height of the bubbles.} 
 \label{fig:accessibility}
 \end{figure*}

\section{Experiencing Spatial Orchestra}

Users were instructed to stay within the virtual fence and to move slowly around the play-space. They were further informed that each bubble encountered would contain cello sounds. Color coded bubbles matched specific notes encouraging users to interact with the bubbles individually or in clusters to compose music (Figure \ref{fig:bubbles}c).

The size of the space, bubble dimensions, number of bubbles, and bubble's surface shaders were meticulously adjusted to ensure an optimal user experience. The goal was to create an environment full enough for creative music composition but not to the extent that users felt a lack of control.

\subsection{User Feedback}

During our university public event, we showcased the Spatial Orchestra, allowing interested participants to sign up and experience it. Over the course of two days, 60 users had the opportunity to engage with the Spatial Orchestra.

While the majority of feedback was positive, some users mentioned that they couldn't easily discern whether they were inside the virtual bubble or not, as the visual cue disappeared once they entered the bubble. To address this, we modified the rendering of the bubble texture to be visible from both inside and outside, providing users with a translucent color cue to indicate their position within the bubble (Figure \ref{fig:boundary}b).

In response to user feedback, we also adjusted the transparency levels of the bubbles to enhance visibility, especially when multiple bubbles overlapped. This modification allowed users to better perceive individual bubble colors and easily identify and mix multiple cello notes in the same area. Experienced musicians told us that if they could see clearly when bubbles stacked in a crowded area, they would be able to play music more effectively (Figure \ref{fig:boundary}a). 

Based on participant comments, we observed that many reported favorable experiences with high levels of engagement and felt immediately able to produce music while interacting with bubbles. Some participants even spent more than 20 minutes exploring and mastering the rhythm of the composition, exceeding our expectations for user engagement.

\subsection{Accessibility and Safety}

For users to understand their location in the play-space and proximity to the center, we included an augmented marker indicating the center. Colorful plants signified the boundary of the play-space and the proximity to the fence, adding a visual cue and safety buffer.

Azure Spatial Anchors precisely aligned augmented objects during user motion~\cite{buck2022azure}. Multiple anchors ensured alignment throughout the experience; if one failed, others maintained the objects' locations. Only one anchor failed in a continuous five-hour test. The participant remained unaware, highlighting the system's reliability. Lastly, to accommodate wheelchair users and children, we also implemented Accessibility Mode, which allowed us to manually adjust the height of the bubbles that float.

% Original Longer
%In order for users to understand their location within the play-space and proximity to the center, we have included an indicator for the center marker. Additionally, the presence of colorful plants serves as a visual cue to signify being outside the play-space or too close to the fence, providing an added layer of safety. During a 60-user demo at a university event held in a heavily overcast environment, there were no safety issues reported. The experience itself does not involve aggressive motion; instead, it promotes a calm interaction between users. We have implemented Azure Spatial Anchor to precisely align augmented objects in response to active user motion. Multiple spatial anchors are utilized to ensure consistent alignment of augmented objects throughout the user experience. In the rare event of one anchor point failing, the other anchors maintain the projected items' locations. During five hours of continuous testing at the university event, only one anchor point experienced a temporary failure, while the other anchor successfully preserved its position. Importantly, the participant remained unaware of this occurrence, highlighting the reliability and durability of this mechanism.

\section{Conclusion}
In conclusion, we introduced Spatial Orchestra, a spatial musical instrument that utilizes natural locomotion. Playing this instrument is not only an immersive experience but also an expression of physical movement. Utilizing a stand-alone augmented reality headset, we showcased a music instrument that leverages walking as a means of interaction. It is designed for anyone to learn quickly and play effortlessly, even without prior musical training.

During a university event, participants discovered various patterns and techniques to play music. They adjusted the rhythm by entering bubbles more or less frequently, showcasing their virtuosity through empirical user tests. Some users also found creative ways to produce specific sounds by combining bubbles or waiting before entering them.

Users enjoyed the distinctive experience of interacting with bubbles to create sounds. Despite the challenges of generating precise notes and rhythms, the simplicity of using natural inputs like walking offers users an accessible and enjoyable way to play music.

\begin{acks}
This work was supported in part by NSF award IIS-2211784 and by UCSB's Media Arts and Technology Program.
\end{acks}

\bibliographystyle{ACM-Reference-Format}
\bibliography{sample-base}


\begin{thebibliography}{22}

%%% the Web address.

\ifx \showCODEN    \undefined \def \showCODEN     #1{\unskip}     \fi
\ifx \showDOI      \undefined \def \showDOI       #1{#1}\fi
\ifx \showISBNx    \undefined \def \showISBNx     #1{\unskip}     \fi
\ifx \showISBNxiii \undefined \def \showISBNxiii  #1{\unskip}     \fi
\ifx \showISSN     \undefined \def \showISSN      #1{\unskip}     \fi
\ifx \showLCCN     \undefined \def \showLCCN      #1{\unskip}     \fi
\ifx \shownote     \undefined \def \shownote      #1{#1}          \fi
\ifx \showarticletitle \undefined \def \showarticletitle #1{#1}   \fi
\ifx \showURL      \undefined \def \showURL       {\relax}        \fi
% invisible to TeX
\providecommand\bibfield[2]{#2}
\providecommand\bibinfo[2]{#2}
\providecommand\natexlab[1]{#1}
\providecommand\showeprint[2][]{arXiv:#2}

\bibitem[Bilbow(2022)]%
        {Bilbow22}
\bibfield{author}{\bibinfo{person}{Sam Bilbow}.} \bibinfo{year}{2022}\natexlab{}.
\newblock \showarticletitle{Evaluating polaris~ - An Audiovisual Augmented Reality Experience Built on Open-Source Hardware and Software}.
\newblock \bibinfo{journal}{\emph{New Interfaces for Musical Expression}}.
\newblock
\urldef\tempurl%
\url{https://doi.org/10.21428/92fbeb44.8abb9ce6}
\showDOI{\tempurl}


\bibitem[Buck et~al\mbox{.}(2022)]%
        {buck2022azure}
\bibfield{author}{\bibinfo{person}{Alex Buck}, \bibinfo{person}{Radford Parker}, {and} \bibinfo{person}{Matt Wojciakowski}.} \bibinfo{year}{2022}\natexlab{}.
\newblock \bibinfo{title}{Azure Spatial Anchors overview}.
\newblock \bibinfo{howpublished}{https://learn.microsoft.com/en-us/azure/spatial-anchors/overview}.
\newblock
\newblock
\shownote{Accessed: October 2022}.


\bibitem[Caramiaux et~al\mbox{.}(2015)]%
        {FormFollowsSound}
\bibfield{author}{\bibinfo{person}{Baptiste Caramiaux}, \bibinfo{person}{Alessandro Altavilla}, \bibinfo{person}{Scott~G. Pobiner}, {and} \bibinfo{person}{Atau Tanaka}.} \bibinfo{year}{2015}\natexlab{}.
\newblock \showarticletitle{Form Follows Sound: Designing Interactions from Sonic Memories}. In \bibinfo{booktitle}{\emph{Proceedings of the 33rd Annual ACM Conference on Human Factors in Computing Systems}} (Seoul, Republic of Korea) \emph{(\bibinfo{series}{CHI '15})}. \bibinfo{publisher}{Association for Computing Machinery}, \bibinfo{address}{New York, NY, USA}, \bibinfo{pages}{3943–3952}.
\newblock
\showISBNx{9781450331456}
\urldef\tempurl%
\url{https://doi.org/10.1145/2702123.2702515}
\showDOI{\tempurl}


\bibitem[Cheok et~al\mbox{.}(2002)]%
        {cheok2002interactive}
\bibfield{author}{\bibinfo{person}{A.D. Cheok}, \bibinfo{person}{Wang Weihua}, \bibinfo{person}{Xubo Yang}, \bibinfo{person}{S. Prince}, \bibinfo{person}{Fong~Siew Wan}, \bibinfo{person}{M. Billinghurst}, {and} \bibinfo{person}{H. Kato}.} \bibinfo{year}{2002}\natexlab{}.
\newblock \showarticletitle{Interactive theatre experience in embodied + wearable mixed reality space}. In \bibinfo{booktitle}{\emph{Proceedings. International Symposium on Mixed and Augmented Reality}}. \bibinfo{pages}{59--317}.
\newblock
\urldef\tempurl%
\url{https://doi.org/10.1109/ISMAR.2002.1115073}
\showDOI{\tempurl}


\bibitem[Escarce~Junior et~al\mbox{.}(2021)]%
        {Junior21}
\bibfield{author}{\bibinfo{person}{M\'{a}rio Escarce~Junior}, \bibinfo{person}{Georgia Rossmann~Martins}, \bibinfo{person}{Leandro Soriano~Marcolino}, {and} \bibinfo{person}{Elisa Rubegni}.} \bibinfo{year}{2021}\natexlab{}.
\newblock \showarticletitle{A Meta-Interactive Compositional Approach That Fosters Musical Emergence through Ludic Expressivity}, Vol.~\bibinfo{volume}{5}. \bibinfo{publisher}{Association for Computing Machinery}, \bibinfo{address}{New York, NY, USA}, \bibinfo{pages}{32}.
\newblock
\urldef\tempurl%
\url{https://doi.org/10.1145/3474689}
\showDOI{\tempurl}


\bibitem[Gehlhaar(1991)]%
        {soundspace91}
\bibfield{author}{\bibinfo{person}{Rolf Gehlhaar}.} \bibinfo{year}{1991}\natexlab{}.
\newblock \showarticletitle{SOUND = SPACE: an interactive musical environment}.
\newblock \bibinfo{journal}{\emph{Contemporary Music Review}} \bibinfo{volume}{6}, \bibinfo{number}{1}.
\newblock
\urldef\tempurl%
\url{https://doi.org/10.1080/07494469100640061}
\showDOI{\tempurl}


\bibitem[Ivanyi et~al\mbox{.}(2023)]%
        {DuoRhythmo}
\bibfield{author}{\bibinfo{person}{Balazs~Andras Ivanyi}, \bibinfo{person}{Truls~Bendik Tjemsland}, \bibinfo{person}{Christian~Vasileios Tsalidis~de Zabala}, \bibinfo{person}{Lilla~Julia Toth}, \bibinfo{person}{Marcus~Alexander Dyrholm}, \bibinfo{person}{Scott~James Naylor}, \bibinfo{person}{Ann Paradiso}, \bibinfo{person}{Dwayne Lamb}, \bibinfo{person}{Jarnail Chudge}, \bibinfo{person}{Ali Adjorlu}, {and} \bibinfo{person}{Stefania Serafin}.} \bibinfo{year}{2023}\natexlab{}.
\newblock \showarticletitle{DuoRhythmo: Design and Remote User Experience Evaluation (UXE) of a Collaborative Accessible Digital Musical Interface (CADMI) for People with ALS (PALS)}. In \bibinfo{booktitle}{\emph{Proceedings of the 2023 CHI Conference on Human Factors in Computing Systems}} (Hamburg, Germany) \emph{(\bibinfo{series}{CHI '23})}. \bibinfo{publisher}{Association for Computing Machinery}, \bibinfo{address}{New York, NY, USA}, Article \bibinfo{articleno}{56}.
\newblock
\showISBNx{9781450394215}
\urldef\tempurl%
\url{https://doi.org/10.1145/3544548.3581285}
\showDOI{\tempurl}


\bibitem[Kim et~al\mbox{.}(2022)]%
        {kim2022investigating}
\bibfield{author}{\bibinfo{person}{You-Jin Kim}, \bibinfo{person}{Radha Kumaran}, \bibinfo{person}{Ehsan Sayyad}, \bibinfo{person}{Anne Milner}, \bibinfo{person}{Tom Bullock}, \bibinfo{person}{Barry Giesbrecht}, {and} \bibinfo{person}{Tobias Hollerer}.} \bibinfo{year}{2022}\natexlab{}.
\newblock \showarticletitle{Investigating Search Among Physical and Virtual Objects Under Different Lighting Conditions}.
\newblock \bibinfo{journal}{\emph{IEEE Transactions on Visualization and Computer Graphics}} (\bibinfo{year}{2022}), \bibinfo{pages}{1--11}.
\newblock
\showISSN{1077-2626}
\urldef\tempurl%
\url{https://doi.org/10.1109/TVCG.2022.3203093}
\showDOI{\tempurl}


\bibitem[Kim et~al\mbox{.}(2023a)]%
        {kim2023dynamic}
\bibfield{author}{\bibinfo{person}{You-Jin Kim}, \bibinfo{person}{Joshua Lu}, {and} \bibinfo{person}{Tobias H\"{o}llerer}.} \bibinfo{year}{2023}\natexlab{a}.
\newblock \showarticletitle{Dynamic Theater: Location-Based Immersive Dance Theater, Investigating User Guidance and Experience}. In \bibinfo{booktitle}{\emph{Proceedings of the 29th ACM Symposium on Virtual Reality Software and Technology}} (Christchurch, New Zealand) \emph{(\bibinfo{series}{VRST '23})}. \bibinfo{publisher}{Association for Computing Machinery}, \bibinfo{address}{New York, NY, USA}, Article \bibinfo{articleno}{27}.
\newblock
\showISBNx{9798400703287}
\urldef\tempurl%
\url{https://doi.org/10.1145/3611659.3615705}
\showDOI{\tempurl}


\bibitem[Kim et~al\mbox{.}(2023b)]%
        {kim2023reality}
\bibfield{author}{\bibinfo{person}{You-Jin Kim}, \bibinfo{person}{Andrew~D. Wilson}, \bibinfo{person}{Jennifer Jacobs}, {and} \bibinfo{person}{Tobias Höllerer}.} \bibinfo{year}{2023}\natexlab{b}.
\newblock \showarticletitle{Reality Distortion Room: A Study of User Locomotion Responses to Spatial Augmented Reality Effects}. In \bibinfo{booktitle}{\emph{2023 IEEE International Symposium on Mixed and Augmented Reality (ISMAR)}}. \bibinfo{pages}{1201--1210}.
\newblock
\urldef\tempurl%
\url{https://doi.org/10.1109/ISMAR59233.2023.00137}
\showDOI{\tempurl}


\bibitem[Kurucz et~al\mbox{.}(2023)]%
        {AuditoryImmersion}
\bibfield{author}{\bibinfo{person}{Peter Kurucz}, \bibinfo{person}{Nilufar Baghaei}, \bibinfo{person}{Stefania Serafin}, {and} \bibinfo{person}{Eve Klein}.} \bibinfo{year}{2023}\natexlab{}.
\newblock \showarticletitle{Enhancing Auditory Immersion in Interactive Virtual Reality Environments}. In \bibinfo{booktitle}{\emph{2023 IEEE International Symposium on Mixed and Augmented Reality Adjunct (ISMAR-Adjunct)}}. \bibinfo{pages}{789--792}.
\newblock
\urldef\tempurl%
\url{https://doi.org/10.1109/ISMAR-Adjunct60411.2023.00174}
\showDOI{\tempurl}


\bibitem[Lee(2021)]%
        {lee21}
\bibfield{author}{\bibinfo{person}{Myungin Lee}.} \bibinfo{year}{2021}\natexlab{}.
\newblock \showarticletitle{Entangled: A Multi-Modal, Multi-User Interactive Instrument in Virtual 3D Space Using the Smartphone for Gesture Control}.
\newblock \bibinfo{journal}{\emph{New Interfaces for Musical Expression}}.
\newblock
\urldef\tempurl%
\url{https://doi.org/10.21428/92fbeb44.eae7c23f}
\showDOI{\tempurl}


\bibitem[Maes(2016)]%
        {maes2016sensorimotor}
\bibfield{author}{\bibinfo{person}{Pieter-Jan Maes}.} \bibinfo{year}{2016}\natexlab{}.
\newblock \showarticletitle{Sensorimotor grounding of musical embodiment and the role of prediction: A review}.
\newblock \bibinfo{journal}{\emph{Frontiers in psychology}}  \bibinfo{volume}{7} (\bibinfo{year}{2016}), \bibinfo{pages}{308}.
\newblock
\urldef\tempurl%
\url{https://doi.org/10.3389/fpsyg.2016.00308}
\showDOI{\tempurl}


\bibitem[Maes et~al\mbox{.}(2014)]%
        {maes2014action}
\bibfield{author}{\bibinfo{person}{Pieter-Jan Maes}, \bibinfo{person}{Marc Leman}, \bibinfo{person}{Caroline Palmer}, {and} \bibinfo{person}{Marcelo~M Wanderley}.} \bibinfo{year}{2014}\natexlab{}.
\newblock \showarticletitle{Action-based effects on music perception}.
\newblock \bibinfo{journal}{\emph{Frontiers in psychology}}  \bibinfo{volume}{4} (\bibinfo{year}{2014}), \bibinfo{pages}{1008}.
\newblock
\urldef\tempurl%
\url{https://doi.org/10.3389/fpsyg.2013.01008}
\showDOI{\tempurl}


\bibitem[Morreale et~al\mbox{.}(2014)]%
        {Morreale14}
\bibfield{author}{\bibinfo{person}{Fabio Morreale}, \bibinfo{person}{Antonella~De Angeli}, \bibinfo{person}{Raul Masu}, \bibinfo{person}{Paolo Rota}, {and} \bibinfo{person}{Nicola Conci}.} \bibinfo{year}{2014}\natexlab{}.
\newblock \showarticletitle{Collaborative creativity: The Music Room}.
\newblock \bibinfo{journal}{\emph{Personal and Ubiquitous Computing}}  \bibinfo{volume}{18}, \bibinfo{pages}{1187--1199}.
\newblock
\urldef\tempurl%
\url{https://doi.org/10.1007/s00779-013-0728-1}
\showDOI{\tempurl}


\bibitem[Rasamimanana et~al\mbox{.}(2012)]%
        {Rasamimanana12}
\bibfield{author}{\bibinfo{person}{Nicolas Rasamimanana}, \bibinfo{person}{Fr\'{e}d\'{e}ric Bevilacqua}, \bibinfo{person}{Julien Bloit}, \bibinfo{person}{Norbert Schnell}, \bibinfo{person}{Emmanuel Fl\'{e}ty}, \bibinfo{person}{Andrea Cera}, \bibinfo{person}{Uros Petrevski}, {and} \bibinfo{person}{Jean-Louis Frechin}.} \bibinfo{year}{2012}\natexlab{}.
\newblock \showarticletitle{The Urban Musical Game: Using Sport Balls as Musical Interfaces}.
\newblock \bibinfo{journal}{\emph{CHI '12 Extended Abstracts}}, \bibinfo{pages}{1027–1030}.
\newblock
\urldef\tempurl%
\url{https://doi.org/10.1145/2212776.2212377}
\showDOI{\tempurl}


\bibitem[Riddershom~Bargum et~al\mbox{.}(2023)]%
        {RiddershomBargum2023}
\bibfield{author}{\bibinfo{person}{Anders Riddershom~Bargum}, \bibinfo{person}{Oddur Ingi~Kristj{\'a}nsson}, \bibinfo{person}{P{\'e}ter Bab{\'o}}, \bibinfo{person}{Rasmus Eske Waage~Nielsen}, \bibinfo{person}{Simon Rostami~Mosen}, {and} \bibinfo{person}{Stefania Serafin}.} \bibinfo{year}{2023}\natexlab{}.
\newblock \bibinfo{booktitle}{\emph{Spatial Audio Mixing in Virtual Reality}}.
\newblock \bibinfo{publisher}{Springer International Publishing}, \bibinfo{address}{Cham}, \bibinfo{pages}{269--302}.
\newblock
\showISBNx{978-3-031-04021-4}
\urldef\tempurl%
\url{https://doi.org/10.1007/978-3-031-04021-4_9}
\showDOI{\tempurl}


\bibitem[Schlagowski et~al\mbox{.}(2023)]%
        {schlagowski2023wish}
\bibfield{author}{\bibinfo{person}{Ruben Schlagowski}, \bibinfo{person}{Dariia Nazarenko}, \bibinfo{person}{Yekta Can}, \bibinfo{person}{Kunal Gupta}, \bibinfo{person}{Silvan Mertes}, \bibinfo{person}{Mark Billinghurst}, {and} \bibinfo{person}{Elisabeth Andr\'{e}}.} \bibinfo{year}{2023}\natexlab{}.
\newblock \showarticletitle{Wish You Were Here: Mental and Physiological Effects of Remote Music Collaboration in Mixed Reality}. In \bibinfo{booktitle}{\emph{Proceedings of the 2023 CHI Conference on Human Factors in Computing Systems}} (Hamburg, Germany) \emph{(\bibinfo{series}{CHI '23})}. \bibinfo{publisher}{Association for Computing Machinery}, \bibinfo{address}{New York, NY, USA}, Article \bibinfo{articleno}{102}.
\newblock
\showISBNx{9781450394215}
\urldef\tempurl%
\url{https://doi.org/10.1145/3544548.3581162}
\showDOI{\tempurl}


\bibitem[Serafin and Adjorlu(2023)]%
        {Earnormous}
\bibfield{author}{\bibinfo{person}{Stefania Serafin} {and} \bibinfo{person}{Ali Adjorlu}.} \bibinfo{year}{2023}\natexlab{}.
\newblock \showarticletitle{Earnormous: An Educational VR Game about How Humans Hear}. In \bibinfo{booktitle}{\emph{Proceedings of the 29th ACM Symposium on Virtual Reality Software and Technology}} (Christchurch, New Zealand) \emph{(\bibinfo{series}{VRST '23})}. \bibinfo{publisher}{Association for Computing Machinery}, \bibinfo{address}{New York, NY, USA}, Article \bibinfo{articleno}{52}.
\newblock
\showISBNx{9798400703287}
\urldef\tempurl%
\url{https://doi.org/10.1145/3611659.3616887}
\showDOI{\tempurl}


\bibitem[Turchet et~al\mbox{.}(2021)]%
        {Luca21}
\bibfield{author}{\bibinfo{person}{Luca Turchet}, \bibinfo{person}{Rob~Hamilton Hamilton}, {and} \bibinfo{person}{Anil Çamci}.} \bibinfo{year}{2021}\natexlab{}.
\newblock \showarticletitle{Music in Extended Realities}.
\newblock \bibinfo{journal}{\emph{IEEE Access}}  \bibinfo{volume}{9}, \bibinfo{pages}{15810--15832}.
\newblock
\urldef\tempurl%
\url{https://doi.org/10.1109/ACCESS.2021.3052931}
\showDOI{\tempurl}


\bibitem[Wang and Martin(2022)]%
        {wang22}
\bibfield{author}{\bibinfo{person}{Yichen Wang} {and} \bibinfo{person}{Charles Martin}.} \bibinfo{year}{2022}\natexlab{}.
\newblock \showarticletitle{Cubing Sound: Designing a NIME for Head-mounted Augmented Reality}.
\newblock \bibinfo{journal}{\emph{New Interfaces for Musical Expression}}.
\newblock
\urldef\tempurl%
\url{https://doi.org/10.21428/92fbeb44.b540aa59}
\showDOI{\tempurl}


\bibitem[Winter(2006)]%
        {winter2005interactive}
\bibfield{author}{\bibinfo{person}{Robert Winter}.} \bibinfo{year}{2006}\natexlab{}.
\newblock \showarticletitle{Interactive music: Compositional techniques for communicating different emotional qualities}.
\newblock \bibinfo{journal}{\emph{Master's dissertation, University of York}} (\bibinfo{year}{2006}).
\newblock


\end{thebibliography}


\end{document}